\documentclass[12pt,tightenlines,preprintnumbers,%
nofootinbib]{revtex4}
\newcommand{\PRE}[1]{{#1}} % Use if preprint style
\usepackage{bm} \usepackage{epsfig}
\usepackage{amsmath}
\usepackage{latexsym}
\usepackage{amssymb}

\begin{document}

\preprint{\preprint{MIT-CTP-3721}}  

\title{
\PRE{\vspace*{1.5in}}
Hadron Systematics and Emergent Diquarks\footnote{Talk by FW at a workshop at Schloss Ringberg, October 2005.  To appear in the Proceedings.  A proper paper on this work is in preparation.}
\PRE{\vspace*{0.3in}}
}

\author{Alexander Selem$^{a,b}$}
\PRE{\vspace*{.5in}}
\email{aselem@berkeley.edu}
\author{Frank Wilczek$^a$}
\email{wilczek@mit.edu}
\PRE{\vspace*{.2in}}
\address{
$^a$ Center for Theoretical Physics, Laboratory for Nuclear
Science, and Department of Physics, Massachusetts Institute of Technology,
Cambridge, Massachusetts 02139, USA \\
$^b$ Dept of Physics, University of California,
366 LeConte Hall,
MC 7300, Berkeley, CA 94720-0730
}
%\date{May 2005}

\begin{abstract} \PRE{\vspace*{.3in}} 
We briefly review a variety of theoretical and phenomenological indications for the probable importance of powerful diquark correlations in hadronic physics.  We demonstrate that the bulk of light hadron spectroscopy can be organized using three simple hypotheses:  Regge-Chew-Frautschi mass formulae, feebleness of spin-orbit forces, and energetic distinctions among a few different diquark configurations.     Those hypotheses can be implemented in a semi-classical model of color flux tubes, extrapolated down from large orbital angular momentum $L$.  We discuss refinements of the model to include the effects of tunneling, mass loading, and internal excitations.    We also discern effects of diquark correlations in observed patterns of baryon decays.   Many predictions and suggestions for further work appear.
\end{abstract}

%\pacs{xx}
%12.60.Jv Supersymmetric models
%04.65.+e Supergravity
%95.35.+d Dark matter
%13.85.-t Hadron-induced high- and super-high-energy interactions 
%             (energy>10 GeV)

\maketitle

\PRE{\newpage}

\section{Introduction}

Quantum chromodynamics (QCD) is established as the fundamental underlying theory of the strong interaction, but its application to low-energy hadron phenomenology is far from a matter of routine deduction.   Different idealizations and approximations, whose connection to the underlying theory ranges from tight to tenuous, are presently used to describe different phenomena.   Examples include the nonrelativistic quark model, bag model, Skyrme model,  large $N$ approaches,  Regge phenomenology, and others, not to mention the vast literature of traditional nuclear physics.    

With the increasing power of numerical simulation (lattice gauge theory) to provide accurate access to some of what QCD predicts, it becomes an attractive program to build models whose concepts are firmly based in the fundamental theory, and whose underlying parameters can be accessed by numerical simulation.  

The concept of diquarks has deep roots in the fundamental theory, and has been invoked to help illuminate several phenomena in the strong interaction.    At a crude level, the idea is that pairs of quarks form bound states, which can be treated as (confined) particles, and used as degrees of freedom in parallel with quarks themselves.   A more sophisticated discussion should be phrased in terms of strong correlations between pairs of quarks in favorable channels, resulting in a significant energy gap to the appearance of unfavorable channels.   

There are simple theoretical arguments that suggest which diquark correlations are favorable.  These arguments also suggest that the effect of such correlations should be most important for the lightest quarks, and that the correlations are subject to frustration.  

There are several observed phenomena that can be illuminated using these concepts, as we shall review presently.  In most of those applications, unfortunately, the diquarks are not well isolated.  This complicates their interpretation.   

The present investigation was initially motivated as an attempt to address that problem, using the following simple idea.  Because diquarks are confined, we cannot study them in perfect isolation.   But it is plausible that baryons with large values of the angular momentum $L$ form extended bar-like structures, with quarks pushed to the extremities by centrifugal forces.   More specifically, a widely accepted -- and, as we shall see, remarkably successful -- way to model large $L$ mesons and baryons envisions separated, rotating quark-antiquark or quark-(2 quark) configurations joined by an electric flux tube, or string.   Thus we can use large-$L$ baryons to study diquark interactions, and specifically to test whether the quantum numbers that theory predicts to be energetically favorable are in fact so, and by how much.

%%%%%%%%%%%%%%%%%

\section{Motivations}

\subsection{``Theory''}

\begin{description}
\item[1. Color Cancellation:] 
Since two separated quarks are each in the $\bf \bar 3$ representation, while they can combine into a single $\bf 3$ representation, the disturbance produced by the color charges of two quarks in empty space can be halved by bringing them together.   Thus, on very general grounds, we should expect an attractive interaction between quarks, when their color state is antisymmetric.   
\item[2. Spin Minimization:]
Refining this consideration, we can consider the spins of the quarks.   Both one-gluon exchange and instanton calculations indicate that the antisymmetric, spin-singlet state is more favorable energetically than the symmetric, spin-triplet state.   We might again expect this on very general grounds, in that the total spin is associated with a color magnetic moment, which again disturbs the vacuum.  To enforce Fermi statistics overall, then, we must put the quarks in antisymmetric $\bf \bar 3$ representation of flavor $SU(3)$.   Note that the spin-dependent splitting, being of magnetic origin, is intrinsically relativistic.  Formally, at lowest order, it is inversely proportional to the masses of the particles involved.   While it would be overly naive, in the context of strong interaction physics, to take that dependence literally -- in particular, its suggestion of a divergence for zero mass -- we should expect that the splitting will be largest for the $u$ and $d$ quarks, smaller when $s$ is involved, and much smaller for the heavier quarks.   We will use the notation $[ud], (ud)$ for the good and bad diquark configurations involving $u$ and $d$ quarks, respectively, and similarly for other flavor combinations.
\item[3. Repulsion:]
Since, according to the preceding considerations, energy can be lowered by enforcing favorable correlations between quark pairs, we should expect that any effect that disrupts the correlations will induce a repulsive force.  The presence of an additional diquark will cause such disruptions, due both to competing interactions and to fermi statistics.   So we should expect repulsive forces {\it between\/} diquarks that already have favorable correlations.   
\item[4. Emergence:]
Similarly, the nearby presence of a spectator quark will frustrate favorable diquark correlations.   We should expect that the full correlation energy will emerge when the effects of frustration are minimized, for example in baryons at large $L$ (as anticipated above), or in baryons containing a heavy quark (whose spin is ineffective).   
\end{description}

\subsection{Phenomenology}

\begin{description}
\item[1. Ground State Spectroscopy:] 
A classic manifestation of energetics that depends on diquark correlations is the $\Sigma-\Lambda$ mass difference.   The $\Lambda$ is isosinglet, so it features the good diquark $[ud]$; while $\Sigma$, being
isotriplet, features the bad diquark $(ud)$.    The splitting
\begin{equation}
\Sigma(1193) - \Lambda(1116) =  76 ~{\rm MeV}
\end{equation}
is consistent in sign with the expectations sketched above.   Replacing the $s$ quark with the $c$ quark we find (averaging over the $c$ spin) the much larger splitting
\begin{equation}
\frac{2}{3} \Sigma_{c}(2520)^{\frac{3}{2}^{+} }+ \frac{1}{3} \Sigma_{c}(2455)^{\frac{1}{2}^{+}} - \Lambda_c(2285)  = 213 ~{\rm MeV}
\end{equation}
Again, this is consistent with expectations: the $c$ quark's spin is much less potent than the $s$ quark's, and frustrates $[ud]$ less.  
%%%%%%%%%%%%%%%%%%%%%%%%%%%%
\item[2. Structure Functions:]
One of the oldest observations in deep inelastic scattering is that the ratio of neutron to proton structure functions approaches $\frac{1}{4}$ in the limit $x\rightarrow 1$
\begin{equation}\label{structureRatio}
\lim_{x\rightarrow 1} \frac{F_{2}^{n}(x)}{F_{2}^{p}(x)}  \rightarrow \frac{1}{4} 
\end{equation}   
{}In terms of the twist-two operator matrix elements used in the formal analysis of deep inelastic scattering, this translates into the statement
\begin{equation}\label{opValence}
\lim_{n\rightarrow \infty} \frac{\langle p| \bar d \gamma_{\mu_1}\overleftrightarrow\nabla_{\mu_2}\cdots \overleftrightarrow\nabla_{\mu_n} d | p \rangle}
{\langle p| \bar u \gamma_{\mu_1}\overleftrightarrow\nabla_{\mu_2}\cdots \overleftrightarrow\nabla_{\mu_n} u | p \rangle} \rightarrow 0
\end{equation}
where spin averaging of forward matrix elements, symmetrization over the $\mu$s, and removal of traces is implicit, and a common tensorial form is factored out, together
with similar equations where operators with strange quarks, gluons, etc.  appear in the numerator.   
Equation (\ref{opValence}) states
that in the valence regime $x\rightarrow 1$, where the struck parton carries all the longitudinal momentum of the proton, that struck parton must be a $u$ quark.   
It implies, by isospin symmetry, the corresponding 
relation for the neutron, namely that in the valence regime within a neutron the parton must be a $d$ quark.  Then the ratio of neutron to proton matrix elements will be governed by the ratio of the squares
of quark charges, namely $\frac{(-\frac{1}{3})^{2}}{(\frac{2}{3})^{2}} = \frac{1}{4}$.  Any (isosinglet) contamination from other sources will contribute equally to numerator and denominator, 
thereby increasing this ratio.   Equation (\ref{opValence}) is, from the point of view of symmetry, a peculiar relation: it requires an emergent conspiracy between isosinglet and isotriplet operators.   It is also, from
a general physical point of view, quite remarkable: it is one of the most direct manifestations of the fractional charge on quarks; and it is a sort of hadron = quark identity, closely related to the
quark-hadron continuity conjectured to arise in high density QCD.   

It is an interesting challenge to derive (\ref{opValence}) from microscopic QCD, and to estimate the rate of approach to 0.   The diquark interpretation, put forward by Feynman, is as follows.  It is an extreme behavior, realized in only a small part of the wave function, for one quark to carry almost all the longitudinal momentum.  To enable it, no energy can be wasted: the remaining degrees of freedom should be small in number, and low in energy.   Thus the proton should boil down to no gluons and the minimal number of quarks, i.e. three, with the two spectator quarks in the lowest-energy configuration, i.e. the good diquark $[ud]$.   The remaining, struck quark is therefore always $u$.  

\item[3. Fragmentation:]
A more adventurous application is to fragmentation.  One might guess that the formation of baryons in fragmentation of an energetic quark or gluon jet could proceed stepwise, through the formation of 
diquarks which then fuse with quarks.   To the extent this is a tunneling-type process, analogous to pair creation in an electric field, induced by the decay of color flux tubes, one might expect that the good 
diquark would be significantly more likely to be produced than the bad diquark.   This would reflect itself in a large $\Lambda/\Sigma$ ratio.   And indeed, data from LEP indicates that the value of this ratio is
about 10 for leading particles (that is, at large $z$).  In the Particle Data Book one also finds an encouraging ratio for total multiplicities in $e^{+}e^{-}$ annihilation: $\Lambda_{c}:\Sigma_{c} = .100\pm .03: .014 \pm .007$; in this
case the $c$ quarks are produced by the initiating current, and we have a pure measure of diquarks.  

\item[4. $\Delta I = \frac{1}{2}$ Rule:]
There are also several indications that diquark correlations have other important dynamical implications.  
The $\Delta I = \frac{1}{2}$ rule in strangeness-changing nonleptonic decays has also been ascribed to attraction in the diquark channel.  The basic operator
$\bar u \gamma_{\mu}(1-\gamma_{5})d \bar s \gamma^{\mu} (1-\gamma_{5}) u$ arising 
from $W$ boson exchange can be analyzed into $\bar{[us]}[ud]$, $\bar{(us)}(ud)$, and related 
color-{\bf{6}} diquark types.   Diquark attraction in $\bar{[us]}[ud]$ means that there is a larger chance for quarks in this channel to tap into short-distance components of hadronic wavefunctions.  
This effect is reflected
in enhancement of this component of the basic operator as it is renormalized toward small momenta.  Such an enhancement is well-known to occur at one-loop order (one gluon exchange).   
\item[5. (Paucity of) Exotics:]
The basic degrees of
freedom in QCD include massless gluons and almost-massless $u,d$ quarks, and the interaction strength, though it ``runs'' to small coupling at large 
momentum transfer, is not uniformly small.   We might therefore anticipate, heuristically, 
that low-energy gluons and quark-antiquark pairs are omnipresent, and in particular that the eigenstates of the Hamiltonian -- hadrons -- will be complicated 
composites, containing an indefinite number of particles.   And indeed, according to the strictest experimental measure of internal structure available, the structure 
functions of deep inelastic scattering, nucleons do contain an infinite number of soft gluons and quark-antiquark pairs (parton distributions $\sim \frac{dx}{x}$ as $x\rightarrow 0$).
Yet the main working assumption of the quark model is that
hadrons are constructed according to two body plans: mesons, consisting of a quark and an antiquark; and baryons, consisting of three quarks.  That paucity of body plans seems to conflict with heuristic expectations.

The puzzles posed by the success of the quark model come into sharp focus in the question of {\it exotics}.   Are there additional body plans in the hadron spectrum, 
beyond $qqq$ baryons and $\bar q q$ mesons (and loose composites thereof)?  If not, why not; if so, where are they?   As a special case: why don't multi-nucleons merge into single bags, e.g. 
$qqqqqq$ -- or can they?   The tension between {\it a priori\/} expectations of complex bound states and successful use of simple models, defines the main problem of 
exotics: Why aren't there more of them?

A heuristic explanation can begin along the following 
lines.   Low-energy
quark-antiquark pairs are indeed abundant inside hadrons, as are low-energy gluons, but they have (almost) vacuum quantum numbers: they are arranged in flavor and spin singlets.  
(The ``almost'' refers to chiral symmetry breaking.)   Deviations from the ``good'' quark-antiquark or gluon-gluon channels, which are color and spin singlets, cost significant energy.  States which contain such
excitations, above the minimum consistent with their quantum numbers,  will tend to be highly unstable.    They might be hard to observe as resonances, or become unbound altogether.

The next-best way for extraneous quarks to appear is, according to the preceding considerations, in the form of  ``good'' diquark pairs.   Thus a threatening (that is, superficially promising) strategy for constructing
low-energy exotics apparently could be based on using those objects as building-blocks.   On reflection, there are two reasons  ``good'' diquark correlations help explain the paucity of exotics.  First: due to their total antisymmetry, good diquarks conceal spin and
flavor.  So even if present, they can be hard to discern.  

Second, and more important: because of their mutual repulsion, they inhibit mergers.
Why do protons and neutrons in close contact retain their integrity, rather than merging into a common bag?  A closely related question arises in a sharp form for the $H$ dibaryon that R. Jaffe studied extensively.   It has the configuration $uuddss$.   
In the bag model it appears that a single bag containing these quarks supports a spin-0 state that is quite favorable energetically.   A calculation based on quasi-free quarks 
residing in a common bag, allowing for 
one-gluon exchange, indicates that $H$ might well be near or even below $\Lambda \Lambda$ threshold, and thus strongly stable; or perhaps even below $\Lambda n$ threshold, and therefore stable even against
lowest-order weak interactions.   These possibilities appear to be ruled out both experimentally and by numerical solution of QCD (though possibly neither case is completely airtight).    Good diquark correlations, together with
repulsion between diquarks, suggests a reason why the almost-independent-particle approach fails in this case.    Note that this mechanism requires essentially nonperturbative 
quark interaction effects, beyond one gluon exchange, since one gluon exchange is favorable for $H$.    

\end{description}

\section{Spectroscopy}

%\subsection{Isolating Diquarks: Large $L$}

%%%%%%%%%%

\subsection{Hypotheses in Spectroscopy}

As previewed above, we want to exploit large $L$ baryons to isolate diquarks dynamically.   Since $L$ is not a direct observable, but rather a parameter appearing in the context of a spectroscopic model, to access $L$ we must specify the hypotheses of our model with sufficient precision to permit mapping between predicted states and observed mesons and baryons.   In practice we did this iteratively: trying out tentative hypotheses, using them to make tentative mappings, and refining or rejecting the hypotheses according to our perception of the overall quality of the fit.   We did not attempt to do sophisticated statistical analysis, which given the the complex and spotty nature of the data set would be extremely challenging and probably futile.   Fortunately the preponderance of the established meson and baryon states can be uniquely fit with appropriate quantum numbers and approximate mass using three simple, physically motivated hypotheses, as follows: 

\begin{description} 
\item[1. Loaded Flux Tube:]
The equation $M^{2}= \sigma L$ relating mass to orbital angular momentum arises from solving the equations for a spinning relativistic string with tension $\sigma/(2\pi)$, terminated by the
boundary condition that both ends move transversely at the speed of light.  We might expect it to hold asymptotically for large $L$ in QCD, when an elongated flux tube appears string-like, the rotation is rapid, quark masses are negligible, and semiclassical quantization of its rotation becomes appropriate.   The famous Chew-Frautschi formula $M^{2} = a +  \sigma L$, with simple non-zero values of $a$
(e.g., $a=\frac{1}{2}\sigma$) can result from quantization of an
elementary non-interacting string, including zero-point energy for string vibrations. 

We can generalize the Chew-Frautschi formula by considering two masses $m_1$, $m_2$ connected by a relativistic string with constant tension, $T$, rotating with angular momentum $L$. 
Our general solution naturally arises in a parameterized form in which the energy,$E$, and $L$ are both expressed in terms of the angular velocity, $\omega$, of the rotating system. 
In the limit that $m_1$, $m_2$ $\rightarrow 0$, the usual Chew-Frautschi relationship $E^2 \propto L$ appears.  

Consider masses $m_1$ and $m_2$ at distances $r_1$ and $r_2$ away from the center of rotation respectively.  The whole system spins with angular velocity $\omega$. 
It is also useful to define:
\begin{equation}
\gamma_i= \frac{1}{\sqrt{1-(\omega r_i)^2}}
\label{eq:gam}
\end{equation}  
where the subscript $i$ can be 1 or 2 (for the mentioned masses). It is straightforward to write the energy of the system:
\begin{equation}
E= m_1\gamma_1 +  m_2\gamma_2 + \frac{T}{\omega}\int_0^{\omega r_1}\frac{1}{\sqrt{1-u^2}}du + \frac{T}{\omega}\int_0^{\omega r_2}\frac{1}{\sqrt{1-u^2}}du.
\label{eq:energy1}
\end{equation}
The last two terms are associated with the energy of the string. Similarly, the angular momentum can be written as: 
\begin{equation}
L= m_1\omega r_1^2\gamma_1 + m_2\omega r_2^2\gamma_2 + \frac{T}{\omega^2}\int_0^{\omega r_1}\frac{u^2}{\sqrt{1-u^2}}du + 
\frac{T}{\omega^2}\int_0^{\omega r_2}\frac{u^2}{\sqrt{1-u^2}}du.
\label{eq:angmom1}
\end{equation}
Carrying out the integrals gives:
\begin{subequations}
\label{eq:EandL}
\begin{eqnarray}
E &=& m_1\gamma_1 + m_2\gamma_2 + \frac{T}{\omega}(\arcsin[\omega r_1] + \arcsin[\omega r_2]), \\
L& =& m_1\omega r_1^2\gamma_1 + m_2\omega r_2^2\gamma_2 \\
 && + \frac{T}{\omega^2}\frac{1}{2} \left( -\omega r_1\sqrt{1-(\omega r_1)^2} + \arcsin[\omega r_1] - \omega r_2\sqrt{1-(\omega r_2)^2} +  \arcsin[\omega r_2]\right) \nonumber 
 \,.
\end{eqnarray}
\end{subequations}
Furthermore the following relationship between the tension and angular acceleration holds for each mass:
\begin{equation}
m_i\omega^2 r_i=\frac{T}{\gamma_i^2}.
\label{eq:forces}
\end{equation}

The limit $L \rightarrow 0$ is singular: it requires $\omega \rightarrow \infty$ and $r \rightarrow 0$, and is quite sensitive to the presence of small quark masses or interactions between the ends.  Thus we do not expect our simple semi-classical approach to be accurate in this sector.    The situation is reminiscent of the relation between vortices and rotons in superfluid He$^4$.

The resulting parametric expressions for $E$ and $L$ are opaque to humans, but easy for a computer to handle.  
For the first corrections at small $m_{1}, m_{2}$ (and $L\ne 0$) we find, after some algebra,
\begin{equation}\label{greatFormula1}
E \approx \sqrt {\sigma L} + \kappa L^{-\frac{1}{4}} \mu^{\frac{3}{2}}
\end{equation}
with 
\begin{equation}\label{greatFormula2}
\kappa \equiv \frac{2}{3} \frac{\pi^{\frac{1}{2}}}{\sigma^{\frac{1}{4}}}
\end{equation}
and 
\begin{equation}\label{greatFormula3}
\mu^{\frac {3}{2}} \equiv  m_1^{\frac {3}{2}} + m_2^{\frac {3}{2}}
\end{equation}
This is a useful expression, since it allows us to extract expressions for quark and diquark mass differences from the observed values of baryon and meson mass differences.  
Numerically, $\kappa \approx 1.15$ GeV$^{-\frac{1}{2}}$ for $\sigma \approx 1.1$ GeV$^{2}$.  

Note that the usual correction ascribed to a zero-point vibrations, i.e. a classic intercept of the type $E^{2} = a + (2\pi T)L $, yields corrections of the form 
$E \rightarrow \sqrt {\sigma L} + \frac{a}{2\sqrt {\sigma L}}$.  It becomes subdominant to mass corrections at large $L$.  

For heavy-light systems the corresponding formula is
\begin{equation}
E-M = \sqrt{\frac{\sigma L}{2}} + 2^{\frac{1}{4}} \kappa L^{-\frac{1}{4}} \mu^{\frac{3}{2}}
\end{equation}
where $M$ is the heavy quark mass and $\mu$ is the light quark mass.  The effective tension is halved.

\item[2. Emergent Diquarks:]
We allow diquarks with different quantum numbers and masses at the end of the flux tubes.  
\item[3. Smallness of Spin-Orbit Forces:]
If one inserts a linear confining potential between quark and antiquark into the Dirac equation, the spin-orbit forces depend on whether the potential is taken as a scalar or the fourth component of a vector.  These two options give answers of equal magnitude but opposite sign.  The magnitude is much too large for the data to accommodate.   We obtain a good fit by simply ignoring spin-orbit forces.   
\end{description}

The nature of the fits is best conveyed by reference to the enclosed Tables.  The next subsection provides a running commentary upon them.    

\subsection{Assignments}

\begin{center}
\includegraphics[width=6in]{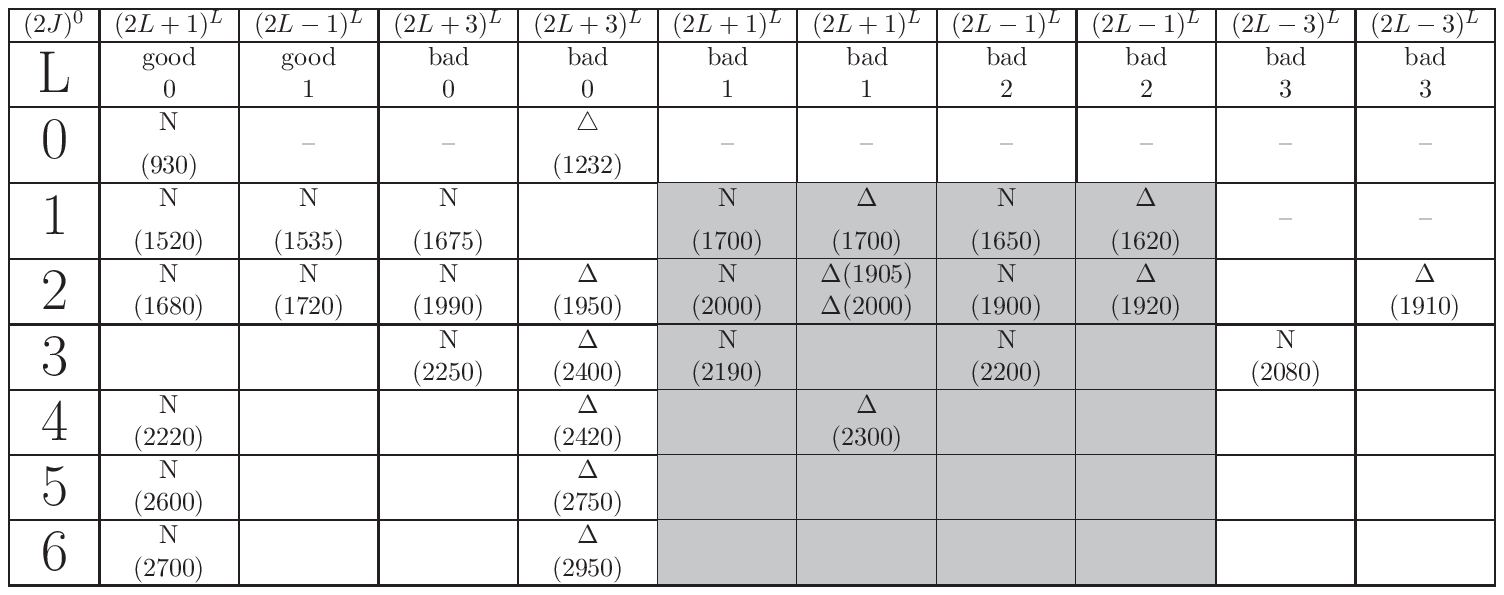}\\
{\small Table 1.  Classification of non-strange baryons.  See text for explanation.}
\label{fig1}\\[1ex]
\end{center}

In this first table, we display assignments for non-strange baryons.   The rules of the game are as follows.   Good diquarks have spin 0 and isospin 0.  Thus when they are assembled into non-strange baryons the total isospin is $\frac{1}{2}$ (i.e., they are nucleons $N$) and the spin-parity $J^P$ is $(L + \frac{1}{2})^L$ or $(L - \frac{1}{2})^L$, each represented once   (except for $L=0$,  of course).  In general, we put a dash $-$ in boxes where no state is expected.   We expect these states to be approximately degenerate, reflecting the smallness of spin-orbit forces.   Bad diquarks have spin 1 and isospin 1.   They can be joined with the other quark into either isospin $\frac{3}{2}$ $\Delta$s or isospin $\frac{1}{2}$ nucleons.   They can also be joined into various spin-partities, as indicated in the Table.  The integer that appears below ``good" or ``bad" in the second row indicates the misalignment, that is how much the total angular momentum differers from its maxiumum for the given spin and orbital angular momenta.   The boxes in gray indicate quantum numbers that can be reached in two ways.    In one case -- $\Delta(1905), \Delta(2000)$ -- two states have been resolved; we predict that better measurements will reveal widespread doubling.  (There are several additional known cases of doublings of this type elsewhere in the baryon and meson spectrum, as we shall see.)   All the ``bad" diquark states in a given row should be approximately degenerate, according to the hypotheses of small coupling between the ends of the flux tubes (including small flavor-dependent interactions) and weak spin-orbit forces.    The $L$ assignments are of course constrained by the Chew--Frautschi-Regge mass formula as well as by quantum numbers.    

The approximate degeneracies our model predicts are well represented in the data, as is the systematic splitting between good and bad diquark configurations, which saturates around 200 MeV at $L=2$.

\begin{center}
\includegraphics[width=6in]{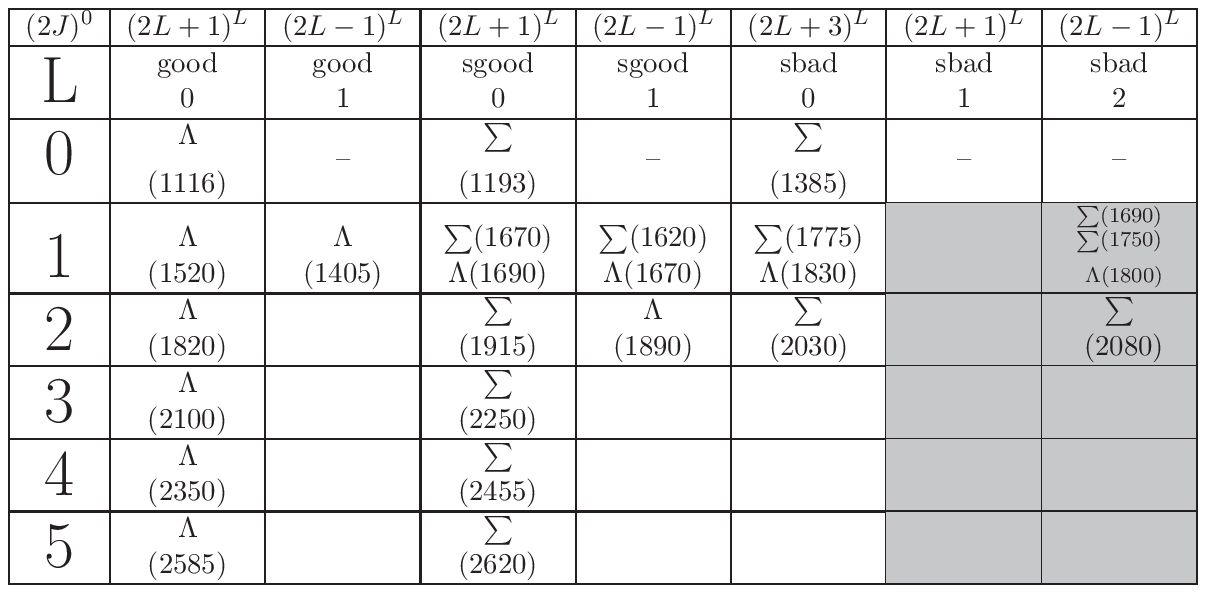}\\
{\small Table 2.  Classification of S=1 baryons.  See text for explanation.}
\label{fig2}\\[1ex]
\end{center}

After the preceding discussion, only a few comments need be added for the other Tables.  The $\Lambda(1405)$ is very light for its position in the table, presumably reflecting the influence of the nearby $KN$ threshold.    The favorable energetics of good versus bad diquark configurations is apparent in the comparison of different columns.  The quantum numbers of columns 4 and 5 are consistent with the body-plan $[us]-d$.    The energy difference between this and the $[ud]-s$ body plan saturates at about 100 MeV.  
We can reach additional spin-parities  with the ``sbad'' body plan $(us)-d$ or with $(ud)-s$.   Only the former supports isospin-$\frac{1}{2}$ $\Lambda$s, so the appearance of near-degenerate $\Lambda$s and $\Sigma$s is evidence for the former body plan (though the $\Sigma$s could be mixtures).   

\begin{center}
\includegraphics[width=3in]{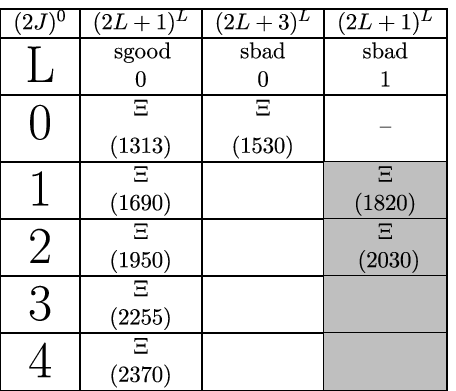}\\
{\small Table 3.  Classification of S=2 baryons.  See text for explanation.}
\label{fig3}\\[1ex]
\end{center}

This sparse table provides some additional support for the diquark distinctions just discussed. 

These baryon tables include most of the relevant resonances classified 2$*$ or better by the Particle Data Book.  The few exceptions will be discussed below (Section IV C).   

\begin{center}
\includegraphics[width=6in]{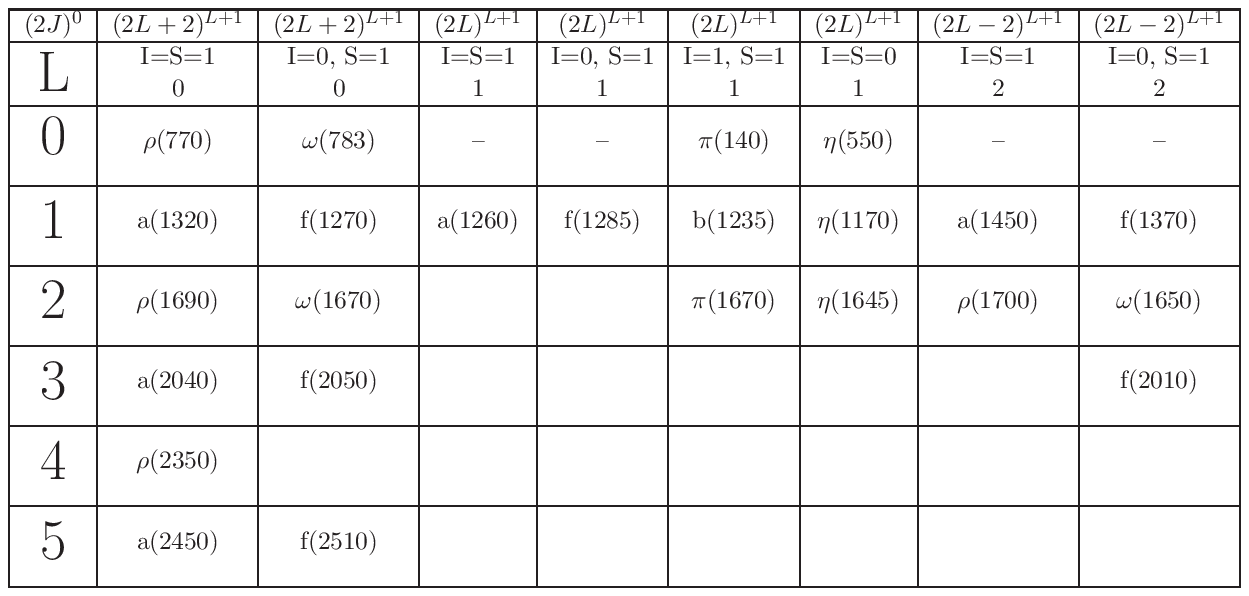}\\
{\small Table 4.  Classification of nonstrange mesons.  See text for explanation.}
\label{fig4}\\[1ex]
\end{center}

The systematic approximate degeneracies across the rows, ranging over different ways of arranging the spins relative to each other and to the orbit, and the family indices, is striking.   It supports the hypotheses of weak spin-orbit forces and dynamical independence between objects on the ends.    Note that the ``doubling'' phenomenon we mentioned as (mostly) an unfulfilled prediction for baryons is here reflected in the many cases where the same spin-parity can be reached in two ways (columns 2/3, 4/5, 6/7, and 8/9).   For mesons separation is much easier, since the two symmetrized eigenstates have different flavor quantum numbers.  

\begin{center}
\includegraphics[width=3in]{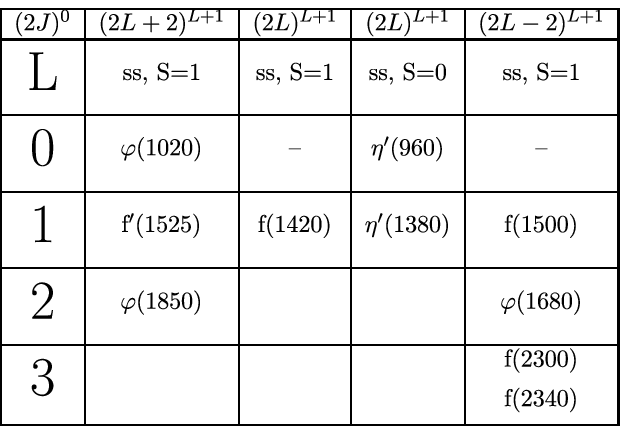}\\
{\small Table 5.  Classification of hidden strangeness mesons.  See text for explanation.}
\label{fig5}\\[1ex]
\end{center}

The $\phi(1680)$ is uncomfortably light.  There are two states $f(2300), f(2340)$ where the $\bar s -s$ body plan allows only one.   The extra state is plausibly ascribed to pure glue (see Section IV C).

\begin{center}
\includegraphics[width=3in]{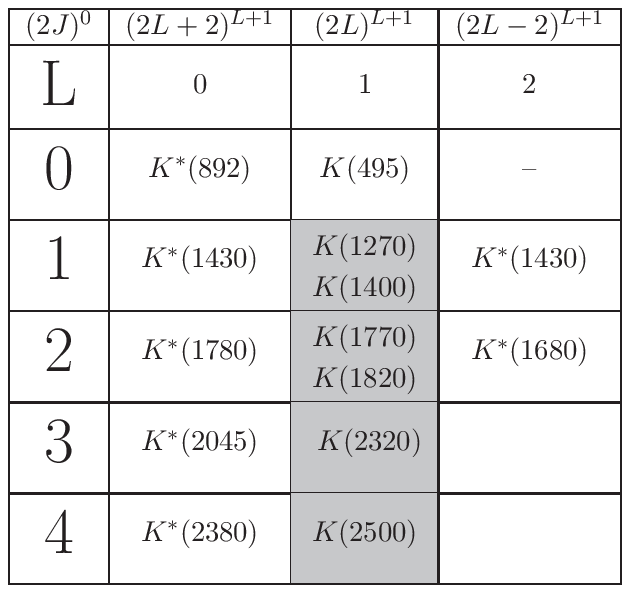}\\
{\small Table 6.  Classification of strange mesons.  See text for explanation.}
\label{fig6}\\[1ex]
\end{center}

Still more approximate degeneracies and doubling.  

These meson tables contain most of the relevant resonances classified awarded a bullet $\bullet$ by the Particle Data Group.  The few exceptions will be discussed below (Section IV C).

\subsection{Regge-Chew-Frautschi Fits}

\begin{center}
\includegraphics[width=5in]{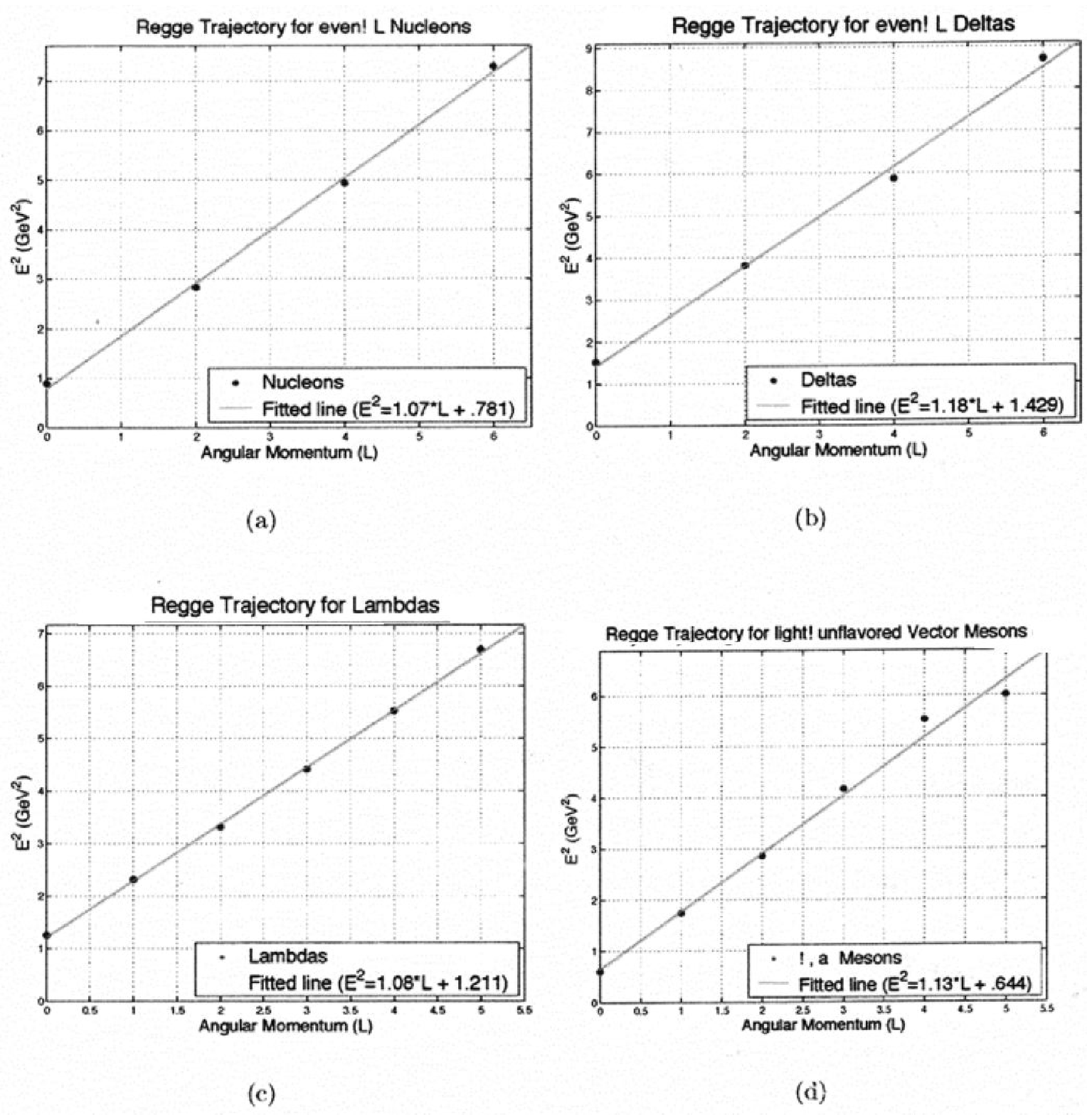}\\
{\small Figure 1.  Fits to some prominent Regge trajectories.  See text for explanation.}
\label{fig2}\\[1ex]
\end{center}

This Figure shows some of the more prominent Regge trajectories.  The near-linearity down to $L=0$, and the universal slope, are evident.  

\begin{center}
\includegraphics[width=5in]{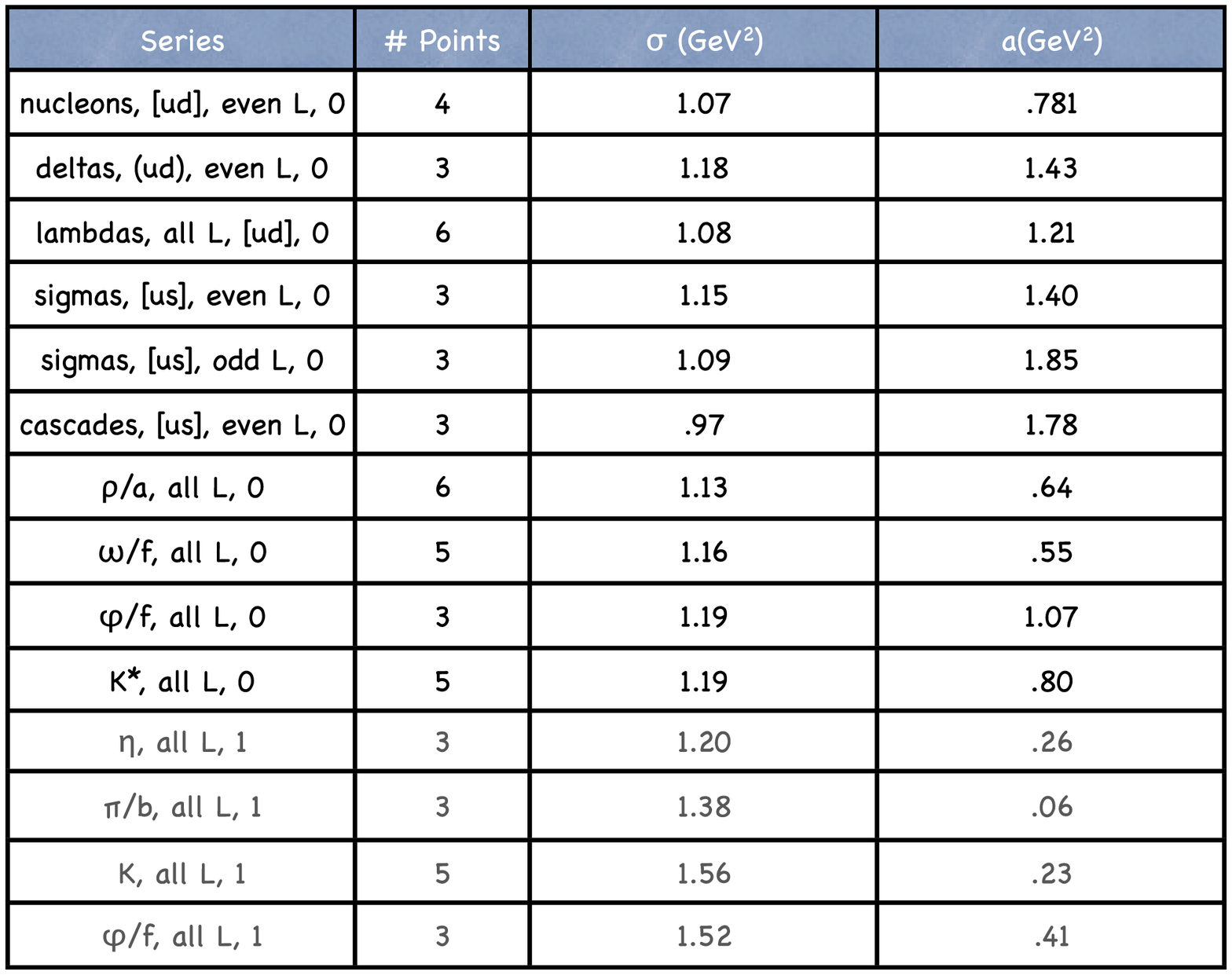}\\
{\small Table 7. Parameters extracted from prominent Regge trajectories.  See text for explanation.}
\label{fig8}\\[1ex]
\end{center}

This Table shows all the trajectories that contain 3 or more points, with their slopes and intercepts.   The bottom four series would appear to favor a larger slope, but in each case there are extenuating circumstances.   The $\eta, \pi/b$ and $K$ series are affected by chiral symmetry breaking.  Their $L=0$ members are approximate Nambu-Goldstone bosons, i.e. largely collective states, and it would be surprising if simple quark-model ideas described them accurately.    Apart from that, the $K(2320)$ is too heavy for its location in the Table: we'd be happy to see it migrate down in mass.   The other bad actor is $\phi(1680)$, mentioned previously.  It infects the bottom row.   

For purposes of this Table we have not attempted to separate mass effects from an intrinsic intercept.

\subsection{The $L=2$ Slice}

\begin{center}
\includegraphics[width=5in]{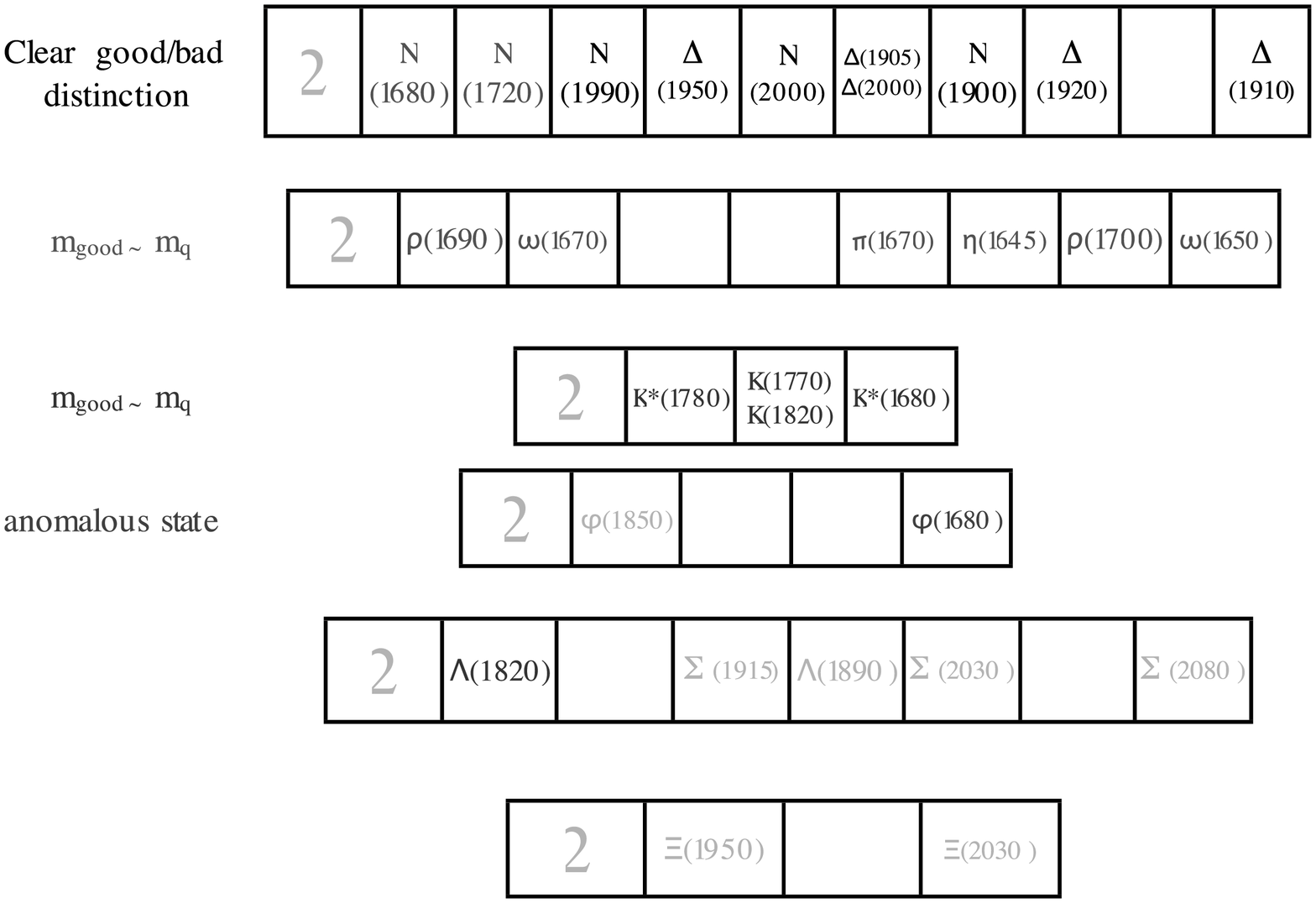}\\
{\small Table 8.  Close-up on the $L=2$ sector.  See text for explanation.}
\label{fig7}\\[1ex]
\end{center}

For low $L$ the hypotheses of our model become dubious, and for very large $L$ the data becomes sparse.  Fortunately, $L=2$ appears to be large enough for simple dynamics to apply, and yet has enough data to make the case powerfully.   

\begin{description}
\item[1. Diquark Distinctions:] 
In the first row, the first two $N$ entries (light gray) contain the good diquark, while the remainder contain the bad diquark.  The two sets are clearly distinguished.   In the fourth row, the good diquark $\Lambda$ is clearly split below the others; the rest come in two tiers, plausibly representing $[us]-d$ and $(us)-d$, as discussed previously.   Similarly for the final column.  
\item[2. Dynamical Independence:]  
The masses of the mesons and baryons are little affected by how we combine the spin and flavor of the two ends.  This, together with the following point, are shown by the near-degeneracy we find among states in each row, once diquark splittings are taken into account. 
\item[3. Feeble Spin-Orbit Forces:]
As just mentioned.
\item[4. Good Diquark - Antiquark Degeneracy:] 
The near-degeneracy between the baryon and meson states in light gray, and between the $\Lambda(1820)$ and $K$ mesons states in dark gray, exhibits the near-degeneracy between the good diquark and a light antiquark. 
\item[5. Anomalous State:]
All this striking success highlights the anomalous nature of the one thing that doesn't fit: $\phi(1680)$ is too light.   
\end{description}

\section{Supplements}

\subsection{Even-Odd Effect}

\begin{center}
\includegraphics[width=4in]{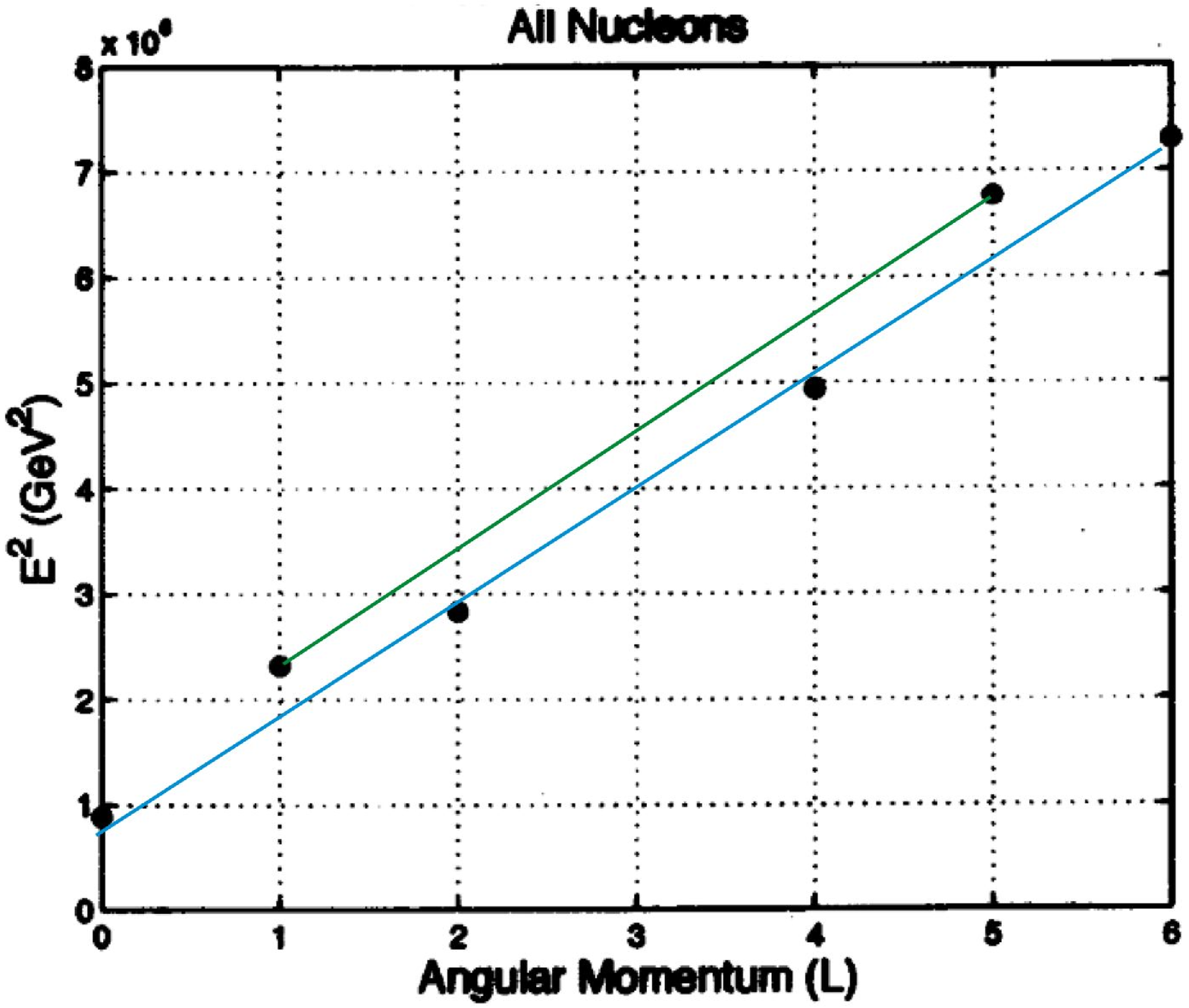}\\
{\small Figure 2.  Splitting of even from odd $L$ trajectories.  See text for explanation.}
\label{angular}\\[1ex]
\end{center}

While the $\Lambda$ trajectory for $M^2$ against $L$ is remarkably straight, passing through all integers from 0 to 5, this is not always the case.  For example, as shown in Figure 2, the 
series of nucleons with different values of $L$, to which we ascribe the body plan $[ud]-d$, do not quite lie on a straight-line trajectories.  Rather, they seem to split into two trajectories, one for even $L$, and a higher one for odd $L$.   There appear to be similar effects in other sectors (though the data is poor).

\begin{center}
\includegraphics[width=2.5in]{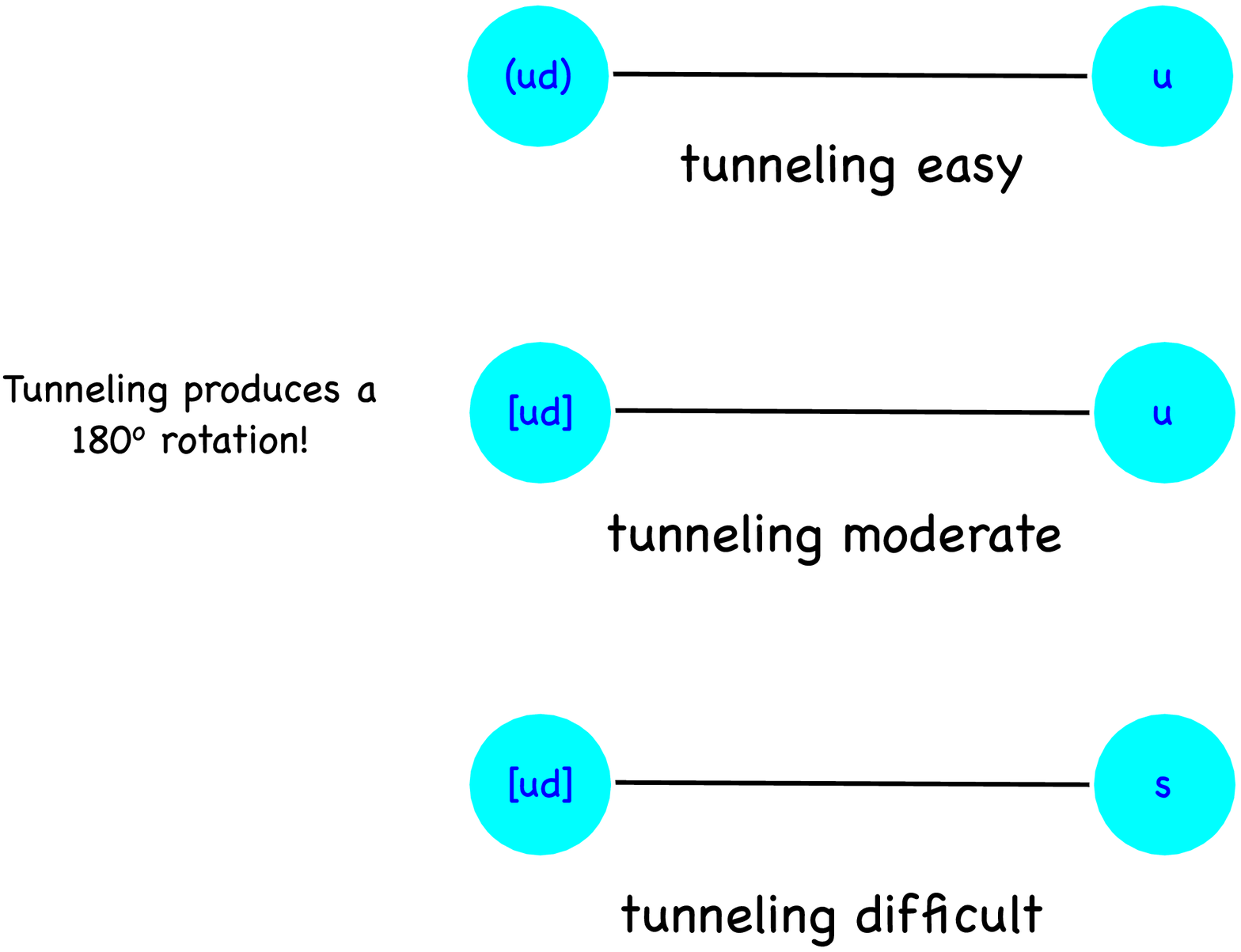}\\
{\small Figure 3.  Possibility of tunneling in different configurations.  See text for explanation.}
\label{tunneling}\\[1ex]
\end{center}

There is a nice explanation for the difference, shown in Figure 3.  Before considering rotation levels, we should fix the internal wave function.   Now if tunneling is possible, we should take symmetric or antisymmetric superpositions between the two permutations of the ends.  Since interchange of the ends has the same effect as a 180$^{\rm o}$ rotation, symmetric wave functions must be associated to even $L$, antisymmetric to odd $L$.  Since a node in the wave function is costly, the odd $L$ should be higher in mass.  This is what we see in the nucleons.   On the other hand, for the good-diquark $\Lambda$ to permute its two ends, all three quarks would have to tunnel.  So in that case the even-odd splitting should be very small, as is observed.

\subsection{Charmed Mesons and Baryons}

We will discuss this subject in more detail elsewhere.   Existing data is sparse, but certainly consistent with the ideas discussed here.  For example, the series $\Lambda_c(2285), \Lambda_c(2625), \Lambda_c(2880)$ at $J^P = \frac{1}{2}^+, \frac{3}{2}^-, \frac{5}{2}^+$ is well fit using the mass-loaded string formulae with $M_c=1600, m_{[ud]} = 180$ MeV$^2$,  and $\sigma = .974$  GeV$^2$.  

\subsection{Additional States}

There are a few resonances that don't fit into our Tables naturally.  We'll briefly discuss those now.  

\begin{description}
\item[1. Glueball Candidates:] 
The $f(1710)$, with $J^{PC} = 0^{++}$ does not fit, and there are two $2^{++}$ states $f(2300), f(2340)$ where only one belongs.  These extra states have the right quantum numbers to be glueballs, and their masses are in broad agreement with expectations from lattice gauge theory.  
\item[2. Daughters?]
$\pi(1300), \eta(1295), \eta(1440)$ are extra $0^{-+}$ states, as is $\pi(1800)$.  $K^*(1410), \rho(1450), \omega(1420)$ are extra vector states.   Formally, they could be accommodated as daughters of the corresponding ground-state mesons, i.e. as internal excitations of the flux tube.  Of course the classical flux tube picture is badly strained at this $L$, but we note with a smile that the masses of these ``daughters'' differ approximately by $\delta M^2 = \sigma$ from the corresponding mesons ($2\sigma$ in the case of $\pi(1800)$).   At this point, we've mentioned all the relevant meson candidates.
\item[3. Pentaquarks?? and Baryon Daughters]
We assume $\theta^+$ is gone for good.   $N(1440), \Lambda(1600), N^*(1710)$ could be daughters of the usual octet, and $\Sigma(1940)^{\frac{3}{2}^+}, \Lambda(2110)^{\frac{5}{2}^-}$ could be $L=1,2$ versions.   There's also a bad diquark analogue of this story, with $\Delta(1600), \Lambda(1810), \Sigma(1880)$ possible $L=0$ daughters and $\Delta(1930)^{\frac{5}{2}^-}, \Delta(1900)^{\frac{1}{2}^-}$ at $L=1$.  At this point, we've mentioned all the relevant baryon candidates.    
\item[4. Diquark-Antidiquark:]
Given the emerging near-degeneracy of $[ud]-q$ and $\bar u -q$ at large $L$, it is hard to resist the inference that $\bar{[ud]} - qq$ states should exist in the same mass range as $u - qq$.   Could there, for example, be a a set of doubly-strange positive parity $J=1,2,3$ mesons with mass close to 2 GeV?
\end{description}

\subsection{Diquarks in Baryon Decays}

Finally, let us mention an interesting dynamical application of our classification.   Since the good $[ud]$ diquark is so favorable, we might expect it to retain its integrity even as the baryon containing it decays.  Specifically, we might expect that $\Lambda$ baryons with the body plan $[ud]-s$ prefer to decay into (generalized) nucleons containing good diquarks and $\bar K$ mesons, as opposed to strange baryons and $\pi$ mesons.   

The $\Lambda(1520)^{\frac{3}{2}^-}$ is a remarkable case.  It decays to $N\bar K$ 45 \% of the time, $\Sigma \pi $ 42 \%, and $\Lambda \pi \pi$ 10 \% (taking the data perhaps too literally).   At first sight these numbers may not appear impressive, but upon reflection they are startling.   Parity and angular momentum conservation bump up the $NK$ channel to d-wave, and it is not very far from threshold.  $\Lambda \pi \pi$ is also phase-space challenged.  So these channels, in which the good diquark retains its integrity, are working against considerable handicaps, yet they hold their own.   

$\Lambda(1820)$ decays 55-65 \% into $N\bar K$, versus 8-14 \% into $\Sigma \pi$.

$\Lambda(2100)$ decays 25-35 \% into $N \bar K$, and another 10-20 \% into $N$ plus excited $\bar K$s, versus 5 \% into $\Sigma \pi$.  

Moving to strange baryons with the body plan $[us]-d$, we expect the opposite effect, that is preference for strange baryon channels with this structure.   Unfortunately the data is very sparse, basically restricting us to $\Sigma(1670)$.   That particle decays 30-60 \% into $\Sigma \pi$ but only 5-15 \% into $\Lambda \pi$ and 7-13 \% into $N\bar K$.  So our expectation is vindicated.  

If we take $[ud]$ integrity at face value, and apply it to our conjectured $\bar {[ud]} - ss$ meson, we are led to expect that $\bar N \Xi$ is a prominent decay channel, if it is kinematically allowed.

\section{Discussion}

Our hope that large $L$ spectroscopy would give convincing evidence for energetically significant diquark correlations is amply fulfilled.  In conjunction with the loaded flux-tube model, assuming negligible spin-orbit forces, this idea forms the basis of a simple and surprisingly successful account of the preponderance of hadron spectroscopy.   

Our hypotheses are quite different in spirit from those adopted in the nonrelativistic quark model, or in any simple potential model.   The observation that pairs of quarks in the good diquark configuration have mass comparable to that of a single quark, which appears quite directly and strikingly in the $L=2$ data, is a qualitative challenge to the foundation of such models.    For relativistic potential or bag models, the smallness of spin-orbit forces poses a qualitative challenge.  

The phenomenological importance of diquarks as building blocks of baryons is difficult to accommodate within large $N_c$ approaches, since diquark phenomenology relies on $N_c = 3 \ngtr  2 + 1$.  Specifically, the approximate degeneracy between mesons and baryons with good diquarks, which is a striking feature of the (moderately) large $L$ data, seems hard to reconcile with the radically different nature of mesons and baryons at large $N_c$.   

Our fits, and the hypotheses that underlie them, suggest that the dynamics of light-quark QCD simplifies at large $L$.   That dynamics appears to be well described by a flux tube or string with universal tension, with quarks or diquarks at the ends.     Elementary ideas about confinement of color suggest that dynamics of this sort should emerge asymptotically at large $L$, but several details are surprising and important.  Semiclassical quantization of rigidly rotating configurations gives a remarkably successful picture of the spectrum down to small $L$, with broad brush even to $L=0$.  There appears to be little interaction between the two objects at the ends, even at $L=1$, and spin-orbit forces are quite small.   Evidence for internal excitations of the flux tube is at best equivocal.   

Challenges to theory: Can these regularities be derived from fundamental QCD?   Can they provide the basis of a systematic approximation scheme?  Can we get good numerical predictions for large $L$ resonances from fundamental theory?  Can we get good numerical data on emergent diquark splittings from fundamental theory?   Apart from spectroscopy, can we make better contact between the dynamical applications of diquark ideas mentioned in ``Motivations'' and fundamental QCD?  

Challenges to experiment: Fill in the Tables, or demonstrate convincingly that they have holes!  Check whether the ``anomalous'' states are real.  Find the predicted diquark-antidiquark states, or rule them out.    Determine whether the heavy-light spectroscopy predicted for mesons and baryons containing $c$ and $b$ together with light quarks matches reality.  

Finally, a common challenge: How important is ``intrinsic noise'' in the hadronic spectrum?   In discerning  our simplicities and broad trends, we did not allow ourselves to be discouraged by occasional discrepancies at the level of 50 or even 100 MeV.   Should we be?  Putting it another way: must we predict that more accurate measurements will bring all the masses into line?  Or does the presence of many nearby states (including continuum states) with degenerate quantum numbers imply some intrinsic scatter into the spectrum, in the spirit of random matrix theory?  If so, can we quantify -- and aspire to predict --  the statistics of the residuals?

\section*{Acknowledgments}

The work of FW is supported in part by funds provided by
the U.S. Department of Energy under cooperative research agreement
DE-FC02-94ER40818.

\end{document}